\documentstyle[12pt,aasms4]{article}
\begin{document}

\def\Mo {$M_{\odot}$}
\def\ein {$e_{1,2}$}
\def\eout {$e_{3}$}
\def\gin {$g_{1,2}$}
\def\Mt {$M_{2}$}
\def\MJ {$M_{J}$}
\def\kms {km s$^{-1}$}
\def\ms {m s$^{-1}$}
\def\Ro {$R_{\odot}$}
\def\rl {$R_{L,2}$}
\def\deg {$^{\circ}$}

\def\oc{third-order }
\def\lot{long-term }

\def\in{_{12}}
\def\out{_3}
\def\einn{e_{1,2}}
\def\eout{e_3}
\def\ain{a_{1,2}}
\def\aout{a_3}
\def\ginn{g_{1,2}}
\def\gout{g_3}
\def\ma{m_1}
\def\mb{m_2}
\def\mc{m_3}
\def\mab{\ma+\mb}
\def\mabc{\mab+\mc}
\def\inc{{\ital i}}
\def\th {\Phi}
\def\thb {{\th}^2}
\def\thba {(1-\thb)}
\def\thca {(3\thb-1)}
    \def\eb {\einn^2}
    \def\sbgin {\sin 2\ginn}
    \def\cbgin {\cos 2\ginn}

\title{THE HIGH ECCENTRICITY\\ OF THE PLANET AROUND 16~CYG~B}

\author{Tsevi Mazeh, Yuval Krymolowski and Gady Rosenfeld}
\affil{School of Physics and Astronomy, Raymond and Beverly Sackler
Faculty of Exact Sciences, Tel Aviv University, Tel Aviv, Israel\\
mazeh@wise7.tau.ac.il}

\vskip 0.8 truecm

{\centerline {Version of Nov. 12th, 1996}}

\begin{abstract}

We consider the high eccentricity, 0.66, of the newly discovered
planet around 16~Cyg~B, using the fact that the parent star is
part of a wide binary. We show
that the high eccentricity of the planet could be the result of tidal
forces exerted on 16~Cyg~B and its planet by 16~Cyg~A, the distant
companion in the system.  By following
stellar triple systems with parameters similar to those of 16 Cyg, we
have established that the orbital eccentricity of the planet could
have gone through strong modulation, with an amplitude of 0.8 or even
larger, with typical timescale of tens of millions years.

The amplitude of the planet eccentricity strongly depends on the
relative inclination between the plane of motion of the planet and
that of the wide binary 16 Cyg AB. To account for the present
eccentricity of the planet we have to assume that the angle between
the two planes of motion is large, at least 60\deg. We argue that this
assumption is reasonable for wide binaries like 16 Cyg AB.

\end{abstract} 
\section {INTRODUCTION}

Very recently, Cochran et al. (1996) have announced the discovery of a
new extra-solar "planet-candidate", with minimum mass of 1.5 Jupiter
masses, orbiting 16 Cyg~B.  The new planet has two special
features. The first one is its orbital eccentricity, 0.66, which is
remarkably larger than the eccentricities of all other seven planets
with a minimum mass smaller than 10 Jupiter masses (e.g. Boss
1996). The other feature is the existence of another star, 16 Cyg~A,
near the parent star of the planet, with a separation of about 40
arcsec.  The proper motion and the radial-velocity of 16 Cyg A and B
(HR 7503 and HR 7504) are very close to each other (Hoffleit \&
Jaschek 1982), so there is no doubt that the two stars are physically
connected.

Both features are very intriguing. Prior to the discovery of 16 Cyg B
planet, Mazeh, Mayor \& Latham (1996) pointed out that the other seven
"planet-candidates" display a correlation between their masses and
orbital eccentricities. Companions with masses smaller than about 5
Jupiter masses have circular orbits while the more massive companions
have eccentric orbits. The new planet, with its low mass and high
eccentricity, breaks this pattern. The other feature, the existence of
the distant companion, 16 Cyg~A, demonstrates for the first time that
a planet with a long period (P=804 days) can be formed in a binary
system.  In this very short note we suggest that the two features of
the new planet {\it might} be correlated.

Mazeh \& Shaham (1979) have shown that in hierarchical stellar triple
systems, the third distant star which orbits the close binary can, in
many cases, modulate the inner binary eccentricity. This comes from
the fact that the simple-minded model of hierarchical triples is only
a first-order approximation. This model assumes that the motion of the
three bodies can be composed of a Keplerian motion of an inner pair
and another Keplerian motion of the center of mass of the inner pair
and the distant star. However, the gravitational attraction of the
tertiary exerted on the two inner bodies is different from the
attraction exerted on a body at the center of mass of the inner binary
system. The difference, sometimes referred to as the tidal force of
the third star, can induce long-term slow modulation of the inner
binary eccentricity. We will show here that the high eccentricity of
the planet around 16~Cyg~B could have resulted from the tidal forces
of 16 Cyg A --- the distant component of the system.

At a distance of 26 pc, the projected separation between 16 Cyg A and B
is about 1000 AU, a lower limit to the actual present separation. The
semi-major axis could be smaller or larger than the present separation,
depending on the eccentricity and the phase of the wide orbit. Somewhat
arbitrarily, we assume here that the semi-major axis of the wide binary
is 1100 AU. The semi-major axis of the planet-candidate around B is 1.7
AU (Cochran et al 1996). We get therefore an hierarchical triple system
with semi-major axis ratio of about 650. This is a relatively large
ratio, considering the fact that some known triples display semi-major
axis ratio of the order of ten (Fekel 1981, Tokovinin 1996).

The large semi-major axis ratio makes the system of 16 Cyg very stable,
and turns the tidal forces of A exerted on B and its planet to be very
small.  One might think therefore that the amplitude of the planet
eccentricity modulation is small. This intuition is wrong. It is true
that the smallness of the tidal forces makes the variation of the planet
eccentricity very slow. However, another effect of the smallness of the
tidal forces of the distant star is to make the period of the
eccentricity modulation very long. Thus, the total increase of the
planet eccentricity, which is being accumulated for a long period of
time, can be very large, and is only weakly sensitive to the semi-major
axis ratio of the system. We will present here simulations which will
demonstrate that in some cases an initial small eccentricity, of, say
0.15, can grow by the tidal forces of the distant star up to 0.8, with a
modulation period of the order of 10 million years.

\section {THE MODULATION OF THE INNER BINARY ECCENTRICITY}

To study the possible configurations which might display a dramatic
modulation of the eccentricity of the planet of Cyg 16 B we treated the
whole system as a triple stellar system, in which the inner binary consists
of 16 Cyg B and its planet, while the tertiary is 16 Cyg A. We
denoted the masses of 16 Cyg B and its planet as $m_1$ and $m_2$,
respectively, while the mass of A will be denoted by $m_3$.

The large ratio of the semi-major axes of the two orbits
makes the period of the inner eccentricity modulation very long.  
Mazeh \& Shaham (1979), using a third-order Hamiltonian of the three-body
problem (Harrington 1968, 1969) estimated the modulation period to be

$$ P_{mod} \sim P_3 \biggl( {P_3\over P_{1,2}} \biggr)
\biggl( {M \over m_3} \biggr) \ , $$
where $P_3$ is the third star orbital period, $P_{1,2}$ is the close
binary period and $M$ is the total mass of the binary system.
Assuming $m_1$, $m_2$ and $m_3$ to be 1, 0.0016 and 1.03 \Mo,
respectively (Cochran et al. 1966), $P_{1,2}$ and $P_3$ to be 2.2 and
25,000 yrs, respectively, we get $P_{mod}$ of the order of
$3\,10^8$ yrs, or 100 million inner orbit revolutions.

To directly integrate Newton equations of the three body problem for 100
million inner revolutions we need days, if not weeks, of CPU time,
even with the fastest available computers. The long CPU time needed
does not allow to search extensively the possible parameter space of
triple system for configurations which display large inner
eccentricity modulation. In particular, we wanted to follow the inner
eccentricity evolution for configurations with different relative angles
between the two orbital motions and different outer eccentricities.
We therefore had to use some approximation to the three body problem
which would enable us to study the characteristics of the inner
eccentricity modulation. Fortunately, such an approximation has been
already developed and was available for following long-term modulation
of triples like the 16 Cyg system.

\subsection { Third-Order Approximation}

The approximation theory we used is based on Harrington's (1968, 1969)
approach, in which the Hamiltonian of the triple system is expanded into an
infinite series with respect to a small parameter --- the ratio of the
semi-major axis of the inner orbit to that of the outer orbit.  Mazeh
\& Shaham (1979) used the third-order Hamiltonian to deduce the
approximate time derivative of $e_{1,2}$, the inner binary eccentricity,
which varies with \gin, the longitude of periastron of the inner
binary. Moreover, Mazeh \& Shaham were able to show that within the
third-order approximation the variation is cyclic and the modulation
is therefore periodic.

Mazeh and Shaham derived explicitly the time derivative of the inner
eccentricity only for small \ein. Following their approach we derive
the full third-order equations to be:

\begin{eqnarray*}
  {d\einn \over dt} & = & {{2\pi} \over {P_{mod}}} \ {15 \over 8} \
              {{\sqrt{1-\eb}} \over {(1-\eout^2)^{3/2}}}
              \thba \einn \sbgin                                 \\
  {d\ginn \over dt} & = & {{2\pi}\over {P_{mod}} } \ {3\over 8} \
{1 \over {(1-\eout^2)^{3/2}}}                                    \\
                   &   &  \times \Biggl[\biggl(G - \th \biggr)
   {{\Phi(2+3\eb-5\eb\cbgin)} \over \sqrt{1-\eb}}
    + \sqrt{1-\eb} \, \biggl( \thca+5\thba\cbgin  \biggr)
    \Biggr]
\end{eqnarray*}
where
$$ G = {{\ma \mb} \over {\mc (\ma + \mb)}}
       \sqrt{ {{\mabc} \over {\mab}} \ 
               {\ain \over \aout}    \
               {{1-\eb} \over {1-\eout^2}}}  \ ,                  $$
$\Phi=\cos\Theta$, and $\ain$ and $\aout$ are the semi-major axes of the
inner and the outer orbit, respectively. These equations are faster to
integrate than Newton equations by a few orders of magnitude.

The ratio of the semi-major axis of the inner orbit to that of the
outer motion for the 16 Cyg system is $1/640$, small enough to suppose
that the third-order approximation does reflect the actual variation
of the inner binary eccentricity.  To make sure this is the case for
the 16 Cyg system, we used a very recent study of the fourth-order
approximation to the three body problem (Krymolowski 1995; Krymolowski
\& Mazeh 1996). For all cases that we checked, with large angle
between the two planes of motion, the fourth-order approximation did
confirm the modulations derived by the third-order equations.

Further, we also compared the third-order approximation of two cases
with direct numerical integration of Newton equations. We ran the code
of Charles Bailyn (1987) who uses the Bulirsch-Stoer integration
scheme. With a precision parameter of $10^{-11}$ per integration step,
we get relative energy error of about $10^{-4}$ for the whole
integration, which included $2\,10^7$ inner revolutions. The code ran
on a Silicon Graphics Power Challenge 10000, with MIPS R10000 running
at 200MHz as CPU. Despite the extremely fast machine, a {\it week} of
CPU time was enough to follow the triple system only for 20 million
inner revolutions. However, for this period of time the result of the
direct numerical integration also confirmed the third-order
approximation for the two cases.

Further, we ran another code which follows a test particle in a binary
system by integrating Newton equations with the Bulirsch-Stoer scheme.
The code followed the test particle for 100 million inner revolutions
and yielded similar results. All these checks convinced us that the
third-order approximation does describe correctly the inner
eccentricity modulation.

In Figure 1 we present two simulations which compare the third-order
approximation with the direct integration.  In both cases, the angle
between the inner and outer planes of motion, $\Theta$, is 60\deg. The
only difference between the two simulations is the outer eccentricity,
$e_3$, which is equal to 0.85 in one case and is equal to 0.60 in the
other case. In both cases the initial value of \gin was 60\deg\ and
that of $g_3$ 10\deg. The third-order approximation was run for 200
million inner revolutions, while the direct integration of Newton
equations goes only for 20 million revolutions.

The aim of bringing these two cases is two folded. We first wish to
show how good is the third-order approximation for the 16 Cyg system
for configurations with high relative inclinations. Indeed, the direct
integration follows exactly the third-order approximation. Secondly,
the figure demonstrates that the inner eccentricity can grow from 0.15
to 0.66 and larger, solely as a result of the tidal force of the third
star.

\subsection { The Amplitude of the Inner Eccentricity Modulation}

Having established the applicability of the third-order approximation
for the 16 Cyg system, we have used this approximation for an
extensive study of the inner eccentricity modulation as a function of
the different parameters of the triple system. It turned out that the
amplitude depends mainly on $\Theta$, as pointed out already by
S\"oderhjelm (1982). To demonstrate this point we followed
S\"oderhjelm and calculated the maximum of the inner eccentricity as a
function of $\Theta$ for the 16 Cyg system. The results of these
calculations are plotted in Figure 2 for $e_3=0.85$. The Figure shows
different curves for different initial conditions of \ein. We see that
the maximum eccentricity rises sharply at about $\Theta=40^{\circ}$,
and that for relative angle close to 90\deg\ we can even get inner
eccentricity close to unity.

One word of caution is due here. For small value of $\Theta$ Figure 2
shows that the maximum inner eccentricity is rather small and
insensitive to $\Theta$. However, this is an artifact of the
third-order approximation. Fourth-order approximation and the direct
integration of newton equations show that in those cases the next term
of the series of expansion of the Hamiltonian is more important and it
contributes substantially to the inner eccentricity modulation
(Krymolowski 1995; Krymolowski \& Mazeh 1996). However, above 40\deg\
the third-order approximation becomes very accurate, as demonstrated
in Figure 1.

\section {Discussion}

By following stellar triple systems with parameters similar to those of
16 Cyg, we have established that the orbital eccentricity of the planet
around 16 Cyg B could have gone through strong modulation because of the
the tidal forces of 16 Cyg A.  The modulation period of the inner
eccentricity of 16~Cyg~B is of the order of 100 million years.  This
period is short relative to the circularization timescale of the planet,
because of its relatively long orbital period. Therefore, the tidal
interaction of 16~Cyg~A can modulate freely the inner eccentricity.

The planet of 16~Cyg~B is probably not the only known planet in wide
binary. It seems as if three other planets, the ones around 55 Cnc,
$\upsilon$ And and $\tau$ Boo, might reside in wide binaries (Marcy,
private communication). However, the planets of those systems have very
short orbital periods, and therefore the circularization timescale is
shorter than the inner eccentricity modulation period. Therefore,
even if the orbital plane of the planet is highly inclined relative to
the wide binary orbit, the circularization processes prevent the inner
eccentricity from growing.

The simulations have shown that the modulation of the eccentricity of
the planet strongly depends on the relative inclination between the two
planes of motion.  Therefore, if the present eccentricity of 0.66 is
indeed the result of the tidal forces of 16 Cyg A, we have to assume
that the angle between the two planes of motion is large, at least
60\deg.

Let us assume that the orbital motion of the planet is aligned with
the stellar rotation of 16 Cyg B. This is very probably the case if
the planet was formed from a disc around 16 Cyg B. Such a disc was
aligned along the angular momentum of the protostar, and so did the
spin of 16 Cyg B. Therefore, the question whether a large angle
between the orbital motions of the planet and the motion of the wide
binary is possible comes down to the question of alignment between the
stellar rotation and the orbital angular momentum of the wide binary.

A study on exactly this issue has been published recently. Hale
(1994) examined a sample of nearby solar-type stars in binaries and checked 
the tendency for alignment between the stellar spin and the orbital angular
momentum. He found such tendency only for binaries with separation
less than 30-40 AU. He did {\it not} find any tendency for binaries with
larger separations. If we accept Hale's findings, we can safely {\it
assume} that the rotational axis of 16 Cyg B is not aligned with the
orbital angular momentum of A and B, and further {\it assume} that the
orbital plane of the planet is also inclined to the the orbital plane
of A and B. The last assumption can explain the high eccentricity of
the "planet candidate".

Is there any observational test, in the present or in the near future,
to confront our suggestion with the observations? We cannot think of
any such test in the foreseen future. The long period of A and B
turns the finding of the exact orientation of the wide binary hopeless
within human lifetime. Thus, there is no hope to measure or estimate
the relative angle between the two motions, even if we measure the
inclination of the planet's orbit relative to our line of sight. The
modulation period of the eccentricity of the planet is even longer, of
the order of 10 million years, and therefore there is no hope to
detect the planet eccentricity variation in the near future. It seems
therefore that the relative angle between the two motions will stay
under the cover of clouds of doubts for a long time.

However, the question whether there exists a correlation between the mass
and the orbital eccentricity of the extra-solar planets probably will
be settled in the near future. The rate of announcements of newly
discovered planets in 1996 is about 0.5 PPM (=Planet~Per~Month). If
this rate will be kept constant for a few years, we will have soon a
large sample of planets, and we will then be able to decide whether
this correlation is a true feature of the extra-solar planets.

We thank Itzchak Goldman and Geoff Marcy for reading the manuscript
and very helpful comments. This work was supported by US-Israel Binational
Science Foundation grant 94-00284.

\newpage

\section*{References}
\begin{description}

\item  Bailyn, C. D. 1987, PhD thesis, Harvard University

\item Boss, A.P. 1996, Physics Today, Sept. 1996, 32

\item Cochran, W.D., Hatzes, A.P., Butler, R.P. \& Marcy, G.W. 1996, Abstract 
of a paper presented at the DPS/AAS October meeting

\item Fekel, F.C. 1981, ApJ, 246, 879

\item Hale, A. 1994, AJ, 107, 306

\item Harrington, R. S. 1968, AJ, 73, 190

\item Harrington, R. S. 1969, Celes. Mech., 1, 200

\item Hoffleit, D. \& Jaschek, C. 1982, Bright Star Catalogue (4th revised
      version) (New Haven: Yale Univ. Observatory)

\item Mazeh, T., Mayor, M. \& Latham, D.W. 1996, ApJ, in press 

%\item Mazeh, T., Krymolowski, Y. \& Latham, D. W., 1993, MNRAS, 263, 775

%\item Mazeh, T. \& Shaham, J., 1976, ApJ, 205, L147

%\item Mazeh, T. \& Shaham, J., 1977, ApJ, 213, L17 

\item Mazeh, T. \& Shaham, J. 1979, A\&A, 77, 145 

%\item S\"oderhjelm S. 1975, A\&A, 42, 229

\item S\"oderhjelm S. 1982, A\&A, 107, 54 

%\item S\"oderhjelm S. 1984, A\&A, 141, 232 

\item Tokovinin, A.A. 1996, A\&A, submitted

\end{description}

\newpage

\figcaption{The inner binary eccentricity as a function of time. The
continuous line is the result of direct integration of Newton
equations, while the crosses are the results of the third-order
approximation. See text for details.}

\figcaption{The maximum of the inner eccentricity as a function of the
relative inclination between the two orbital motion. The outer
eccentricity is 0.85.}

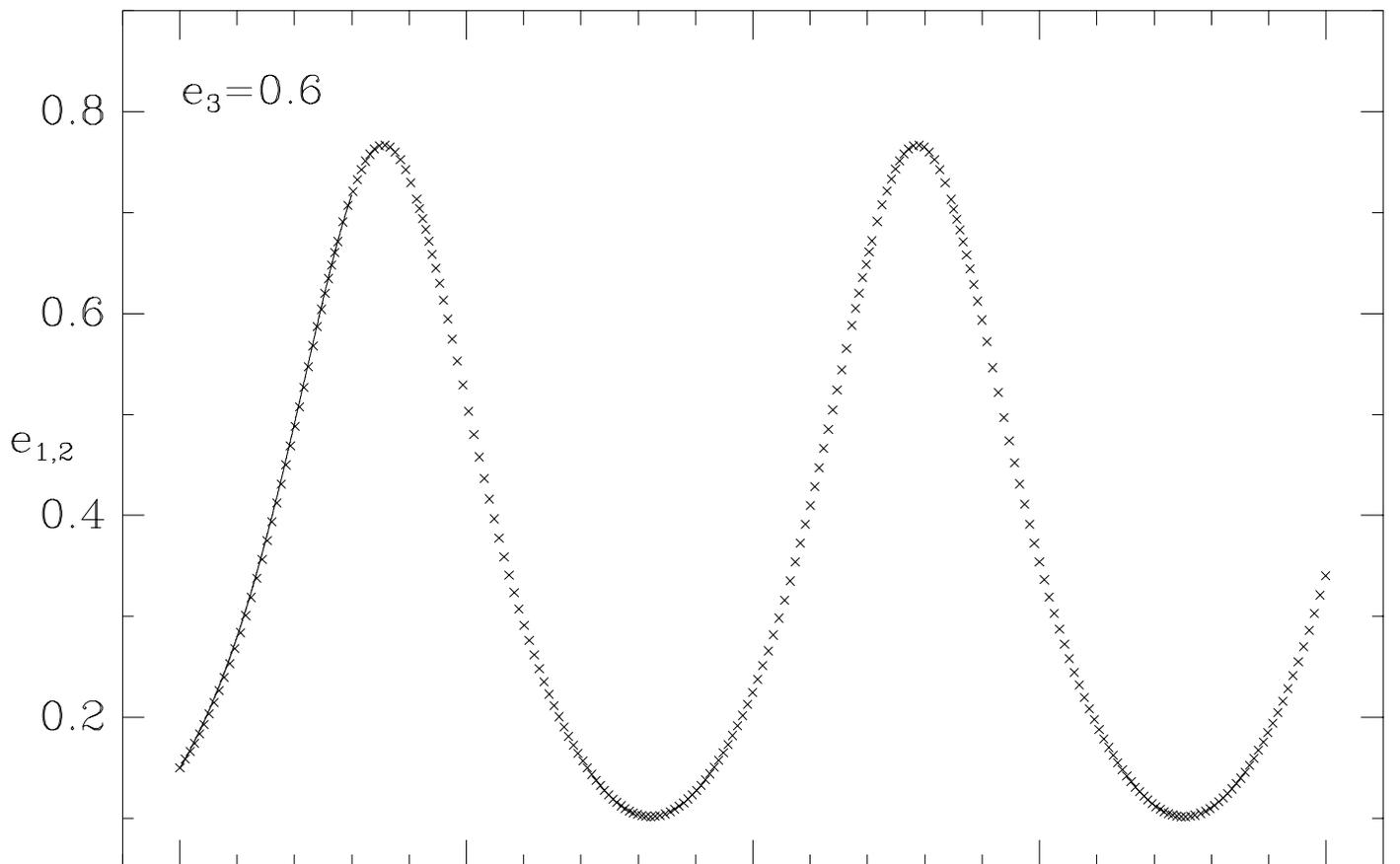
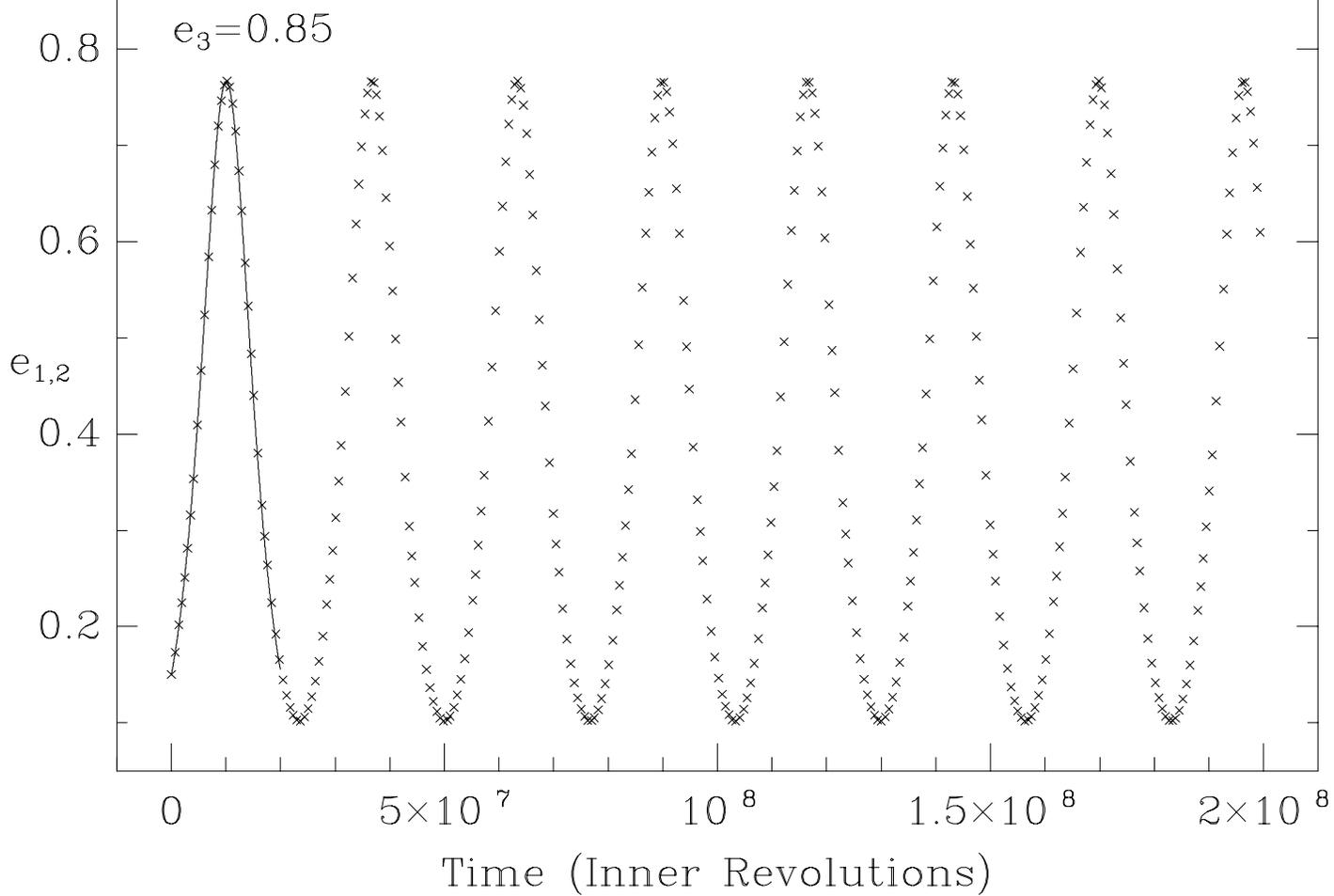
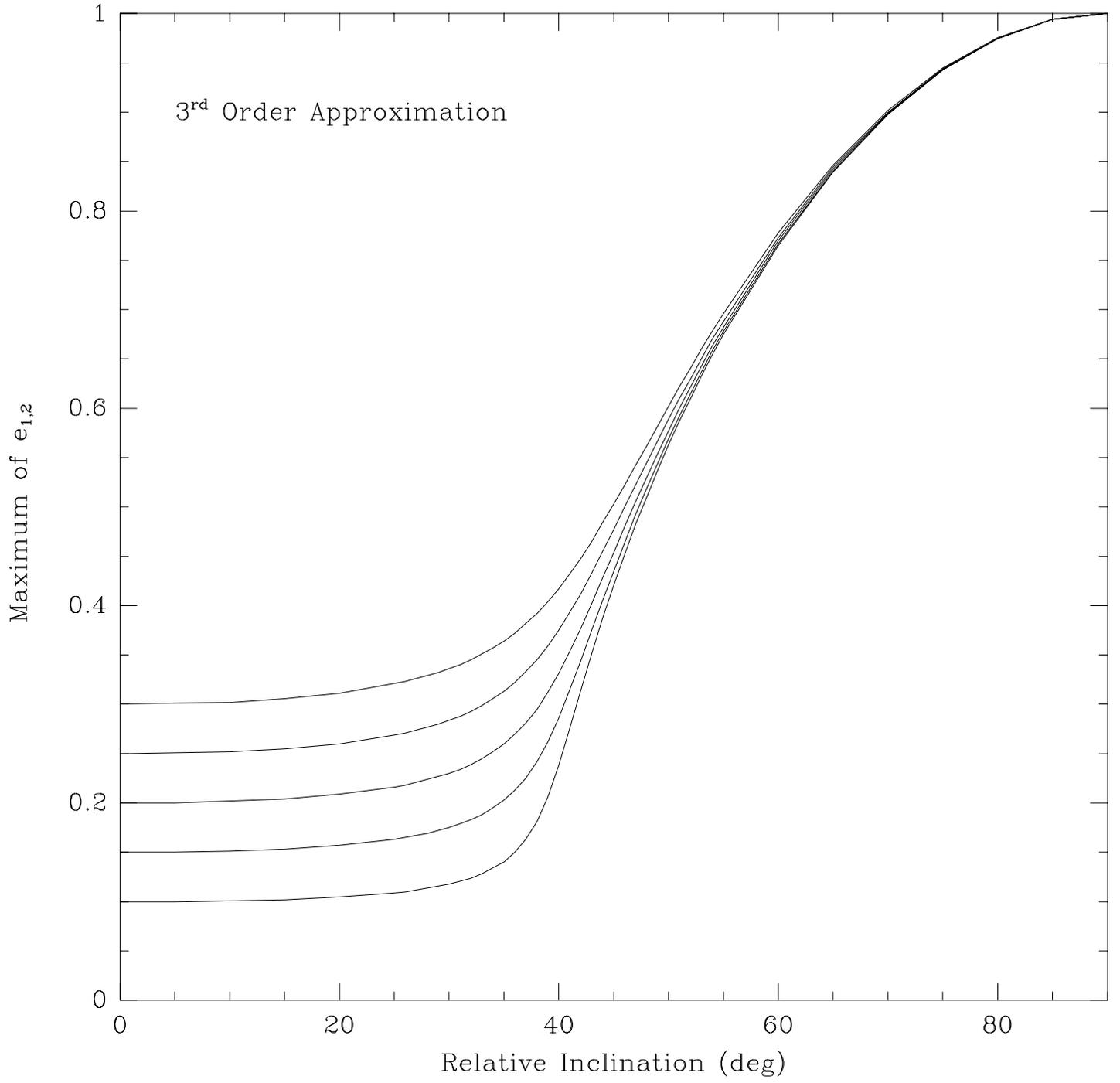
\end{document}